\documentclass[aps,twocolumn,superscriptaddress,floatfix,longbibliography]{revtex4-2}
\usepackage{amsmath,amssymb,amsthm}
\usepackage{physics}
\usepackage{amsfonts}
\usepackage{mathrsfs}
\usepackage{graphicx}
\usepackage{tabularx}
\usepackage{enumerate}
\usepackage{dcolumn}
\usepackage{bm}
\usepackage{xcolor}
\usepackage[normalem]{ulem}
\usepackage[colorlinks,linkcolor=blue,citecolor=blue,urlcolor=blue]{hyperref}

\begin{document}
\title{Exact mobility rings in non-Hermitian quasiperiodically decorated Lieb lattices}

\author{Ming-Jie Tao}
\email{taomingjie1020@sina.com}
\affiliation{College of Mathematics and Physics, Chengdu University of Technology, Chengdu 610059, China}

\author{Yi-Ting Wang}
\affiliation{College of Mathematics and Physics, Chengdu University of Technology, Chengdu 610059, China}

\author{Jing Li}
\affiliation{College of Mathematics and Physics, Chengdu University of Technology, Chengdu 610059, China}



\author{Hongsheng Hou}
\affiliation{School of Physics, Hangzhou Normal University, Hangzhou, Zhejiang 311121, China}


\author{Xiang-Ping Jiang}
\email{2015iopjxp@gmail.com}
\affiliation{School of Physics, Hangzhou Normal University, Hangzhou, Zhejiang 311121, China}

\author{Lei Pan}
\email{panlei@nankai.edu.cn}
\affiliation{School of Physics, Nankai University, Tianjin 300071, China}


\date{\today}

\begin{abstract}
The mobility ring (MR), a critical boundary in the complex energy plane separating extended and localized states, is fundamental to understanding the Anderson transition in non-Hermitian (NH) disordered systems. While MRs have been extensively studied in one-dimensional (1D) NH quasiperiodic models, rigorous analytical frameworks beyond 1D remain critically scarce. Here, we investigate a class of two-dimensional (2D) quasiperiodically decorated Lieb lattices (QDLLs) featuring complex incommensurate potentials selectively applied to the lattice vertices. By exactly mapping these 2D structures onto NH generalized Aubry-Andr{\'e}-Harper (AAH) models and leveraging extended-localized transition point, we analytically derive the Lyapunov exponents and obtain exact  expressions for the MRs. These exact theoretical boundaries are strongly corroborated by numerical computations of wavefunction fractal dimensions and real-space probability distributions. Furthermore, we reveal distinct evolutionary behaviors of the MRs driven by the quasiperiodic potential strength: systems characterized by $\kappa=2$ possess a single MR, whereas systems with $\kappa=3$ undergo a dynamic sequential evolution from a single integrated ring into two independent rings. We hope that our exact results of MRs in 2D will benefit the study of Anderson localizations and MRs in high-dimensional NH systems.  
\end{abstract}

\maketitle

\section{Introduction}\label{sec:1}

Since the seminal discovery of wave-function localization in disordered systems by Anderson in 1958, the Anderson transition (AT) and the mobility edge (ME), which the critical energy boundary separating extended from localized states, have served as central concepts in aperiodic physics~\cite{anderson1958absence,thouless1974electrons,abrahams1979scaling,lee1985disordered,kramer1993localization}. In conventional systems with random disorder, even an infinitesimally small disorder parameter induces global localization of all eigenstates in one-dimensional (1D) and two-dimensional (2D) systems. Conversely, three-dimensional (3D) systems exhibit a genuine Anderson transition governed by the disorder strength, where the ME mediates a continuous metal-insulator phase transition~\cite{evers2008anderson}. As a fundamental bridge between perfectly periodic and completely random limits, 1D quasiperiodic systems, most notably the Aubry-Andr\'{e}-Harper (AAH) model ~\cite{harper1955single,aubry1980analyticity} and its generalizations, have been rigorously proven to host both ATs and distinct MEs~\cite{sarma1988mobility,biddle2010pre,ganeshan2015nearest,luschen2018single,liu2018mobility,wang2020one,liu2022anomalous,wang2023exact,lee2023critical1,zhou2023exact,li2023observation,qi2023multiple,hu2025hidden,wang2025family,li2026multifractal,zhou2026fundamental}.

In the broader context of localization theory, 2D systems are of profound fundamental and practical importance, representing the marginal dimension for the localization transition in real quantum devices~\cite{lee1981anderson,punnoose2005metal,schwartz2007transport}. However, exploring exact analytical solutions in 2D and higher-dimensional systems is notoriously challenging. The severe divergence of the localization length near critical points often leads to significant finite-size effects and convergence bottlenecks in numerical simulations~\cite{bordia2017probing,white2020observation,szabo2020mixed,gautier2021strongly,vstrkalj2022coexistence,cheng2023topological,yang2024exploring}. Furthermore, because the matrix elements of high-dimensional transfer matrices are exceptionally intricate, the classical analytical methods that are highly successful in 1D---such as standard self-duality transformations---cannot be straightforwardly generalized to 2D networks. Consequently, discovering novel lattice geometries and developing innovative theoretical frameworks to analytically derive exact MEs in 2D systems remains a critical bottleneck in contemporary localization physics~\cite{xia2022exact,wang2023two,duncan2023critical,li2024asymmetric,wang2024exact,jiang2024exact,jiang2025localization}.

Meanwhile, the non-Hermitian (NH) Hamiltonian provides an alternative theoretical framework for simulating open quantum systems affected by the coupling of gain, loss, and dissipation. Non-Hermitian systems exhibit an array of exotic physical phenomena without any Hermitian analogues, such as exceptional points~\cite{kawabata2019symmetry,bergholtz2021exceptional,ding2022non,okuma2023non}, the spontaneous breaking of parity-time ($\mathcal{PT}$) symmetry~\cite{longhi2019topological,longhi2022non1,zeng2020topological1}, and the non-Hermitian skin effect~\cite{yao2018edge,yokomizo2019non,song2019non,lee2019anatomy,zhang2020correspondence,okuma2020topological,zeng2020topological2,longhi2022self,lin2023topological,chen2026interplay}. When non-Hermiticity is coupled with localization dynamics, the real energy spectrum naturally extends into the complex energy plane~\cite{liu2020non,liu2020generalized,zeng2020winding,wang2022topological}. In this complex domain, the conventional concept of an ME defined on the real axis generalizes into a closed boundary known as a mobility ring (MR)~\cite{li2024ring,wang2024non,jiang2024exact1,wang2025exact1,pang2025exact,chen2025mobility,he2026exact,wang2026analytical}. Characterizing the exact geometry of these MRs is imperative for comprehensively understanding the transport properties and phase transitions in disordered dissipative systems. Despite extensive investigations into MRs within 1D NH quasiperiodic models, research concerning ATs and MRs in 2D NH systems is still in its infancy. In particular, 2D NH models that are exactly solvable and possess mathematically rigorous mobility boundaries are exceptionally scarce. 

In this work,to bridge this critical gap, we systematically propose and investigate a class of 2D quasiperiodically decorated Lieb lattices (QDLLs) subjected to complex quasiperiodic potentials exclusively at the lattice vertices. By exploiting the inherent multi-sublattice symmetries of this structure, we rigorously map the 2D tight-binding Hamiltonian onto an effective 1D generalized AAH model. Building upon this exact mapping, we employ the dual transformation to analytically calculate the Lyapunov exponents (LEs), thereby deriving exact, closed-form algebraic expressions for the MRs in the complex energy plane. Our analytical and numerical investigations demonstrate that because the LE depends simultaneously on both the real and imaginary parts of the eigenvalues, the NH mobility boundary naturally forms a well-defined MR, perfectly segregating extended states (enclosed within the ring) from localized states (outside the ring). To rigorously validate this universal framework, we examine exact solvable cases and compute the wavefunction fractal dimension, defined via the inverse participation ratio (IPR) as $\eta = -\lim_{N \to \infty} \ln(\text{IPR})/\ln N$, alongside the real-space probability distributions. The numerical simulations exhibit remarkable agreement with our analytically derived boundaries. Furthermore, by tuning the strength of the quasiperiodic potential $V$, we uncover a novel, topology-dependent evolutionary behavior of the MR geometry governed by the modulation parameter $\kappa$. For systems with $\kappa=2$, the complex plane features a single, unified MR. Strikingly, for $\kappa=3$, the MR undergoes a dynamic and sequential topological evolution, splitting from a single integrated ring into two independent, decoupled sub-rings. This work provides a rare and rigorous paradigm for exactly solvable localization transitions in 2D NH systems, establishing a solid theoretical foundation for future explorations of Anderson transitions in higher-dimensional open systems.

The remainder of this paper is organized as follows: In Sec.~\ref{sec:2}, we introduce the 2D VDLL model and detail its exact algebraic mapping to the generalized AAH model. In Sec.~\ref{sec:3}, we apply Avila's global theory to derive the exact MR equations in the complex energy plane. Section~\ref{sec:4} presents the comprehensive numerical verification based on fractal dimensions and discusses the topological evolution of the MRs for $\kappa=2$ and $\kappa=3$. Finally, in Sec.~\ref{sec:5} we summarize our main results and offer a future outlook.

\begin{figure}[t!]
	\centering
	\includegraphics[width=0.48\textwidth]{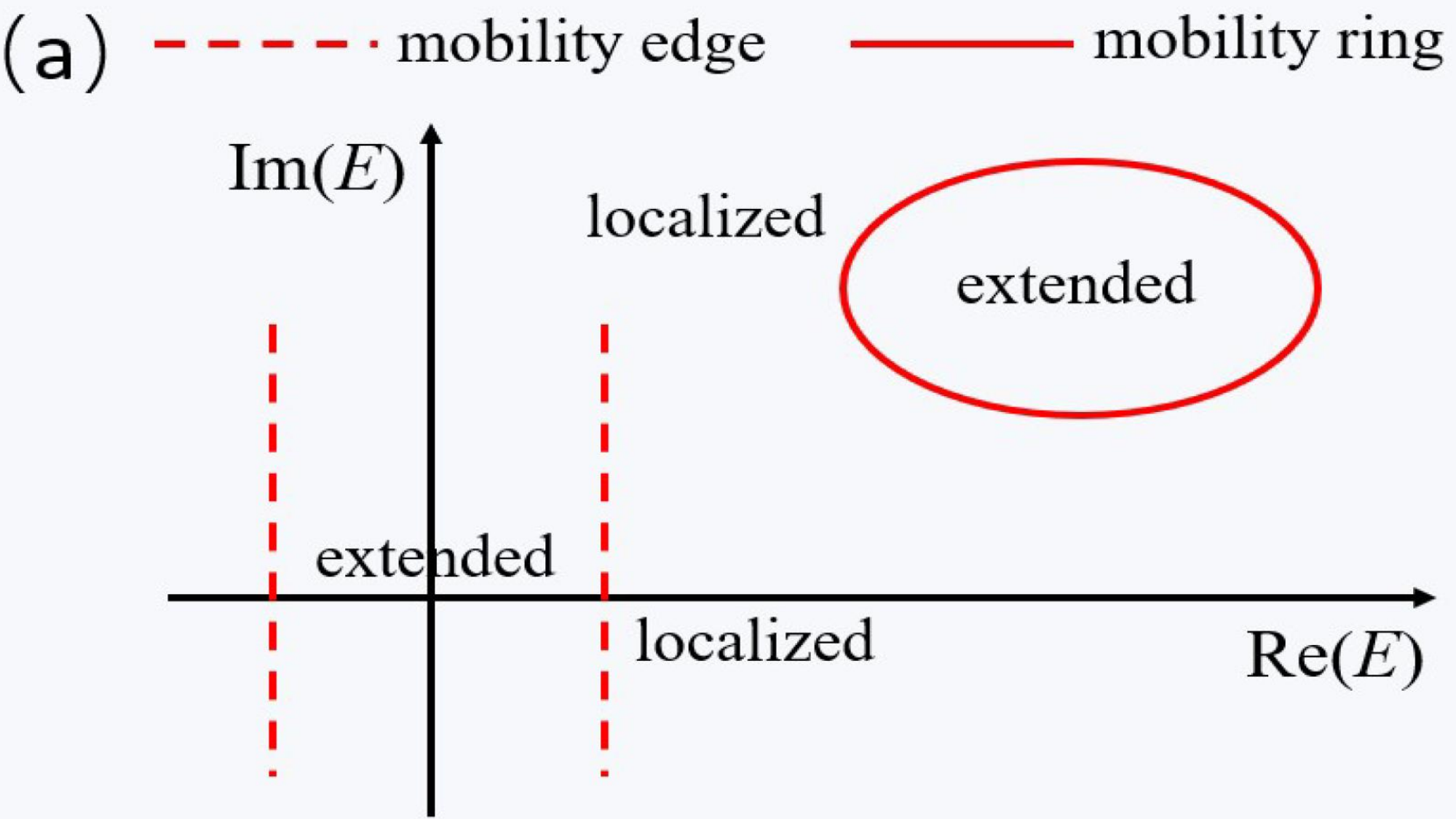}
    \includegraphics[width=0.48\textwidth]{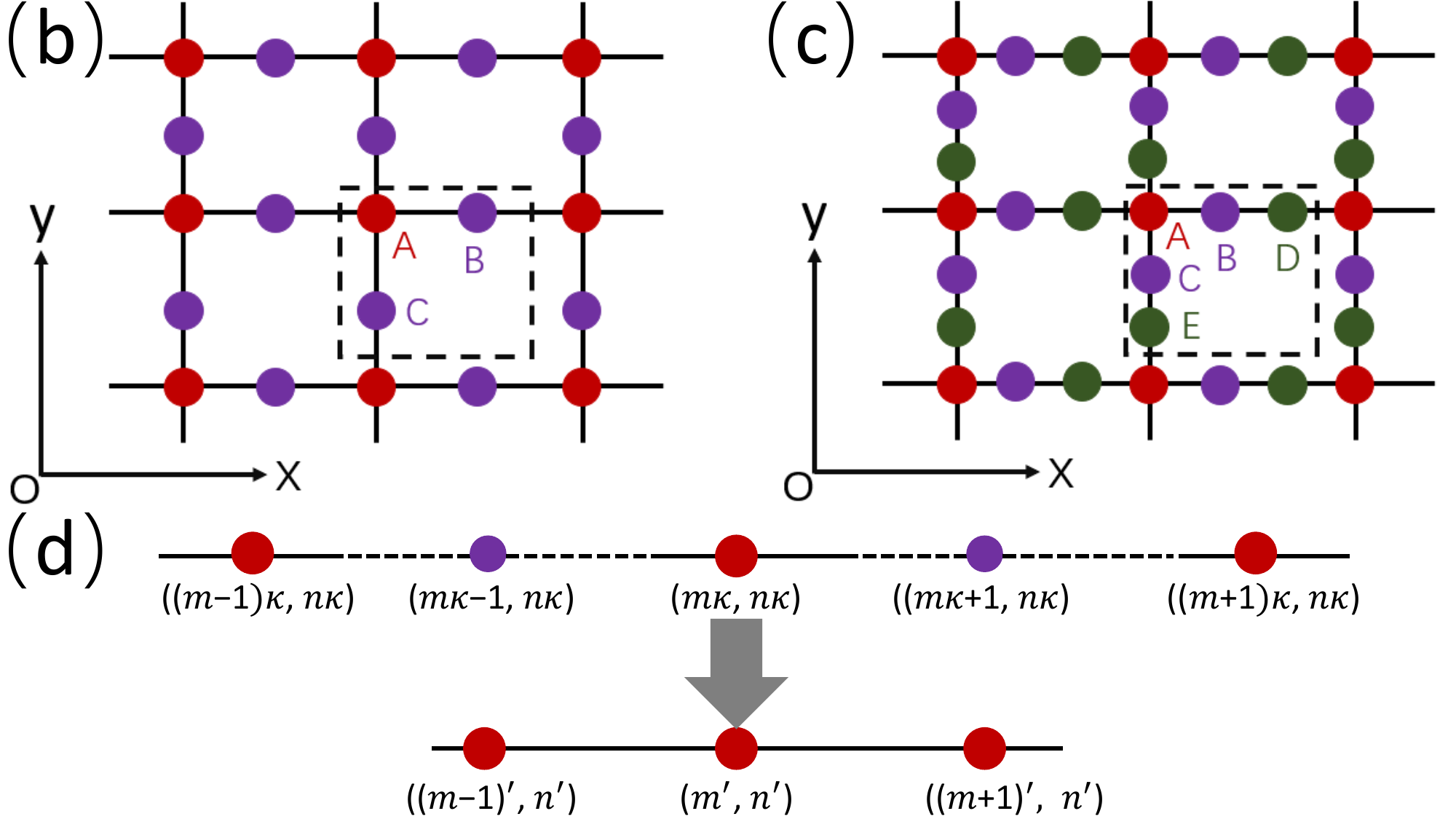}
    \caption{Schematic illustration of the MEs, MRs and the 2D QDLL model. (a) Conceptual phase diagram in the complex energy plane. The solid red curve denotes the MR, which forms a closed boundary perfectly separating the extended states (interior) from the localized states (exterior). The dashed red lines illustrate the conventional MEs projected onto the real energy axis. (b) and (c) Structural representations of the standard 2D Lieb lattice and the extended Lieb lattice, comprising three (labeled A, B, and C) and five (labeled A through E) sub-lattices per unit cell, respectively. The complex quasiperiodic potentials are exclusively introduced at the vertex sites, highlighted as red spheres. 
    (d) Schematic of the mapping from the 2D QDLL model onto an effective AAH model. Here we show the mapping in the $x$ direction, which is similar in the $y$ direction. }
    \label{fig1}
\end{figure}

\section{The model Hamiltonian}\label{sec:2}

We systematically investigate 2D quasiperiodically decorated Lieb lattices (QDLLs) and its extended configurations, wherein complex-valued incommensurate potentials is only acting on the vertices of these lattices. The tight-binding Hamiltonian governing these models is formulated as:
\begin{equation}\label{eq1}
H=\sum_{\langle kl;k'l'\rangle}t(c^{\dagger}_{kl}c_{k'l'}+c^{\dagger}_{k'l'}c_{kl})
+\sum_{kl}V_{kl}c^{\dagger}_{kl}c_{kl},
\end{equation}
where the spatially modulated, complex quasicrystal potential $V_{kl}$ is explicitly defined by:
\begin{equation}\label{eq2}
    V_{kl}=\left\{\begin{matrix} 	2Ve^{i\theta}[\cos(2\pi\alpha_x k+\phi_x)
    \\ \quad\quad+\cos(2\pi\alpha_y l+\phi_y)],\quad & (k,l)=(m\kappa,n\kappa), \\ 	
    0 , \quad & \textrm{otherwise}. \end{matrix}\right.
\end{equation}
Here, $c^{\dagger}_{kl}$ ($c_{kl}$) denotes the creation (annihilation) operator for a spinless fermion occupying the discrete lattice site $(k,l)$. The $t$ represents the hopping amplitude between nearest-neighbor sites. The non-Hermiticity and quasiperiodicity are simultaneously encoded within the on-site potential term $V_{kl}$. Here, the dimensionless parameter $V$ controls the overall modulation amplitude, the complex phase factor $\theta$ breaks Hermiticity of the lattice Hamiltonian and belongs to the range $0$ to $\pi/2$., while the two incommensurate irrational numbers $\alpha_x$ and $\alpha_y$ set the quasiperiodic spatial modulation along the two orthogonal lattice directions. The integer variables $m$ and $n$ index the primitive unit cells along the $x$- and $y$-directions, respectively. We define $L_x$ and $L_y$ as the total number of unit cells comprising the system in the respective directions prior to the introduction of the quasiperiodic modulation. By initializing the spatial coordinates $k$ and $l$ from $0$, we mathematically enforce that the quasiperiodic potential is strictly localized at the geometric vertices—depicted as red spheres in Figs.~\ref{fig1}(b) and \ref{fig1}(c). The geometric frustration inherent to these Lieb lattices architecture has established it as a highly versatile platform for simulating a plethora of non-trivial quantum phenomena. We first establish the geometric and conceptual framework in Fig.~\ref{fig1}(a). Unlike conventional Hermitian systems where the MEs reside strictly on the real energy axis [dashed red lines in Fig.~\ref{fig1}(a)], the non-Hermitian nature of our system drives the spectrum into the complex energy plane. Consequently, the boundary segregating the extended and localized states naturally expands into a closed loop, which denote as the MR [solid red curve in Fig.~\ref{fig1}(a)]. In this work, we aim to study the Anderson localization and exact MRs based on these 2D QDLL models featuring complex incommensurate potentials selectively applied to the lattice vertices. For computational consistency and without sacrificing physical generality, we establish the hopping amplitude $t = 1$ as the unit energy throughout this paper. Furthermore, we fix the incommensurate modulation frequencies to the inverse golden ratio, $\alpha_x = \alpha_y = (\sqrt{5}-1)/2$, and set the arbitrary spatial phase shifts to zero ($\phi_x = \phi_y = 0$). Unless explicitly specified otherwise, all numerical diagonalization are performed under periodic boundary conditions (PBC) to mitigate finite-size boundary effects.

\begin{figure*}[t]
	\centering
	\includegraphics[width=0.80\textwidth]{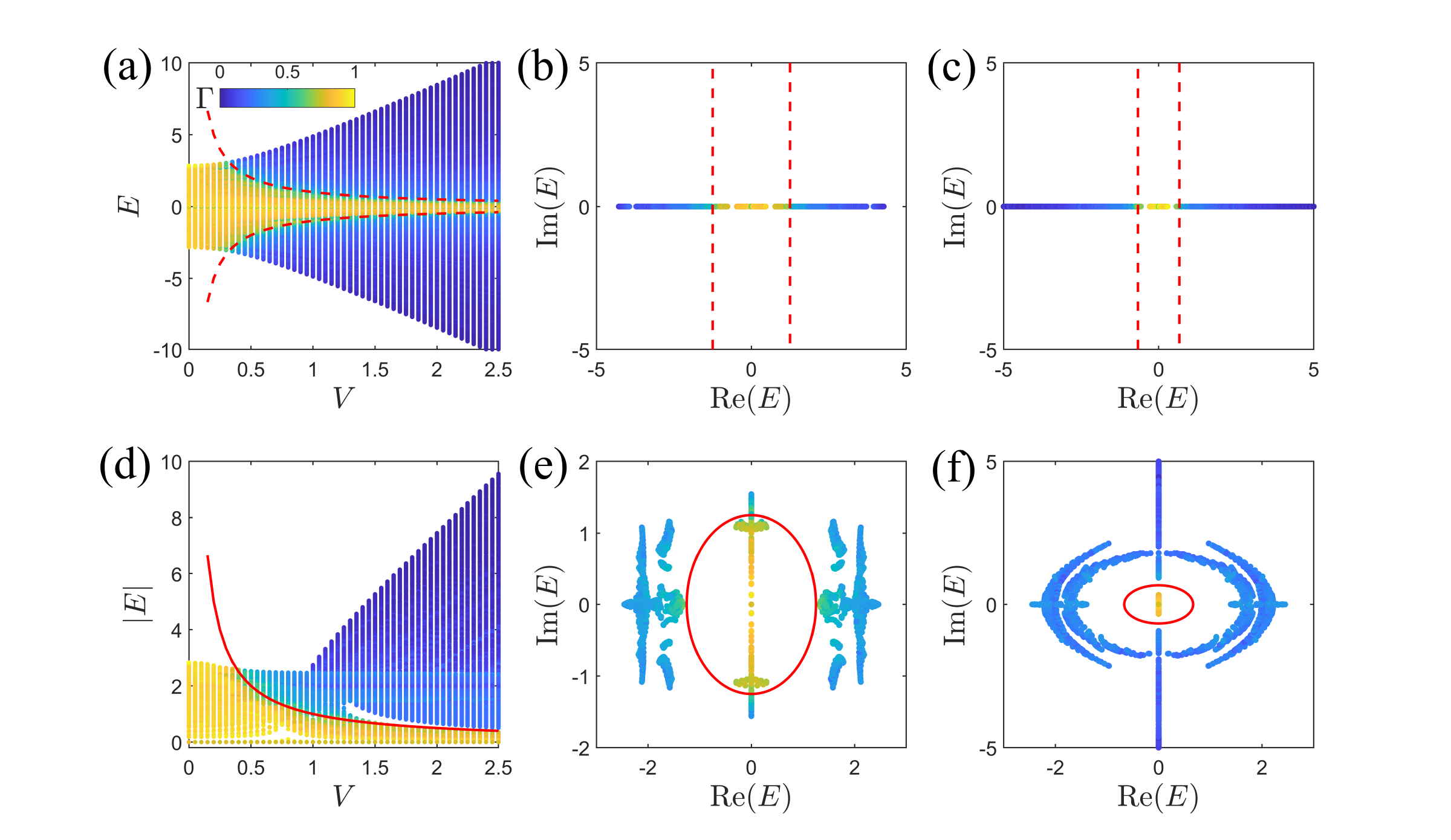}
        \caption{Energy spectrum, mobility edges (MEs), and mobility rings (MRs) for the $\kappa = 2$ case. (a) Wavefunction fractal dimension $\Gamma$ as a function of the quasiperiodic potential strength $V$ and real eigenvalues $E$ in the Hermitian limit ($\theta = 0$). The red dashed lines denote the critical ME boundary $|E_c| = 1/V$, which separating from extended to localized states. (b), (c) Spectral distributions for $V = 0.8$ and $V = 1.5$ at $\theta = 0$. (d) Complex energy spectrum and exact MR boundary $|E| = 1/V$ (solid red curve) for the non-Hermitian case $\theta = \pi/2$. (e), (f) Spectral phase diagrams in the complex energy plane for $V = 0.8$ and $V = 1.5$ at $\theta = \pi/2$. The eigenstates within the MR are extended ($\Gamma \approx 1$), while the eigenstates outside the MR are localized ($\Gamma \approx 0$). Here, $t = 1$ is set as the unit of energy, and the system size are $L_x = L_y = 55$ under periodic boundary conditions.}
	\label{fig2}
\end{figure*}

\begin{figure}[ht!]
	\centering
	\includegraphics[width=0.50\textwidth]{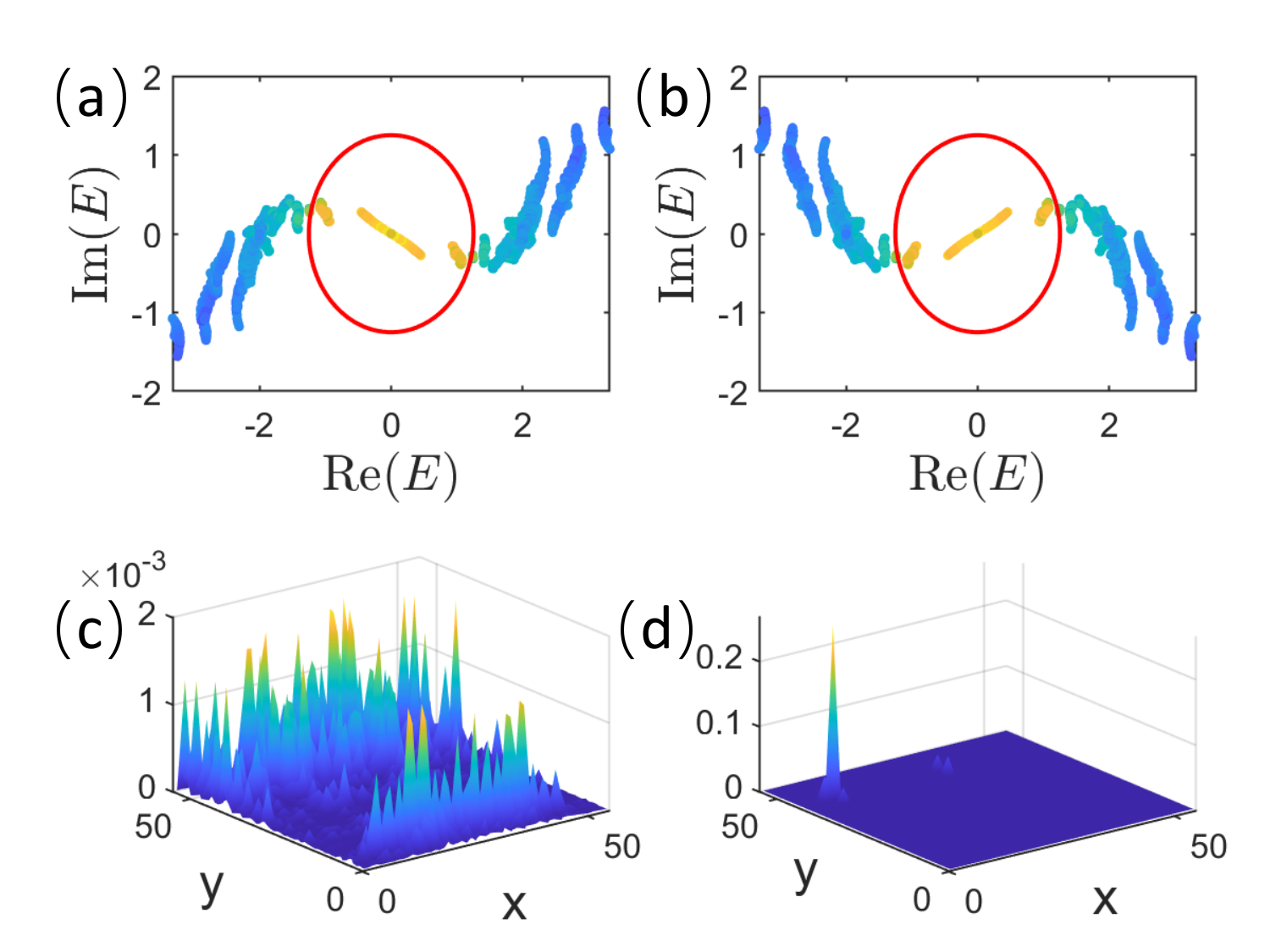}
        \caption{The energy spectrum phase diagram and typical eigenstates distribution. (a)-(b) represent the  $0.75\pi$ Fractal dimension $\eta$ as a function of the quasiperiodic potential strength $V$ and eigenvalues for (b) $\kappa=2$ and (c) $\kappa=3$ with the system size $L_x=L_y=34$.  The spatial distributions of two typical eigenstates with the corresponding complex eigenvalues (c) The complex energy $E=-0.9750 - 0.2198i$ (extended states) and (d) $E=3.2132 - 1.4185i$ (localized states), which are respectively below and above the MR ($|E_c| = 1/V$) of the QDLL model with $V = 1.0$. Here the other parameters are the same as Fig.1.}
	\label{fig3}
\end{figure}

\section{Mapping to the 2D AAH model}\label{sec:3}

To analytically precisely determine localization transition and the the exact MR, we exploit the topological structure of the 2D QDLL to perform an exact algebraic decimation. This procedure maps the original multi-sublattice tight-binding Hamiltonian (\ref{eq1}) onto an effective 2D generalized AAH model. Let us express a general single-particle eigenstate of the system in the real-space tight-binding basis as $|\Psi\rangle = \sum_{k,l} \psi_{k,l} c_{k,l}^{\dagger} |0\rangle$, where $\psi_{k,l}$ is the probability amplitude at the generic lattice site $(k,l)$. Projecting the stationary Schrödinger equation $H|\Psi\rangle = E|\Psi\rangle$ onto the vertex site positioned at $(m\kappa, n\kappa)$ yields the following coupled discrete difference equation:
\begin{equation}\label{eq3}
\begin{aligned}
    E \psi_{m\kappa, n\kappa} = & \psi_{m\kappa+1, n\kappa} + \psi_{m\kappa-1, n\kappa} + \psi_{m\kappa, n\kappa+1} \\
    & + \psi_{m\kappa, n\kappa-1} + V_{m\kappa, n\kappa} \psi_{m\kappa, n\kappa}.
\end{aligned}
\end{equation}
The amplitudes on the undecorated intermediate sites, such as $\psi_{m\kappa+1, n\kappa}$, must be algebraically eliminated to formulate an effective equation solely governed by the vertex amplitudes. By exploiting the translational invariance of the homogeneous hopping along the interstitial chains, we relate the adjacent vertex amplitudes via the standard transfer matrix formalism:
\begin{equation}\label{eq4}
    \begin{pmatrix} \psi_{(m+1)\kappa, n\kappa} \\ \psi_{(m+1)\kappa-1, n\kappa} \end{pmatrix} 
    = T^{\kappa-1} 
    \begin{pmatrix} \psi_{m\kappa+1, n\kappa} \\ \psi_{m\kappa, n\kappa} \end{pmatrix},
\end{equation}
where the elementary transfer matrix for a single unmodulated hopping step is $T = \begin{pmatrix} E & -1 \\ 1 & 0 \end{pmatrix}$. Raising this matrix to the power of $\kappa-1$ yields:
\begin{equation}\label{eq5}
T^{\kappa-1}=\left(
\begin{array}{cc}
  E& -1 \\
  1 & 0 \\
\end{array}
\right)^{\kappa-1} =\left(
\begin{array}{cc}
  a_{\kappa}& -a_{\kappa-1} \\
 a_{\kappa-1} & -a_{\kappa-2} \\
\end{array}
\right),
\end{equation}
where the sequence $a_{\kappa}$ corresponds to the Chebyshev polynomials of the second kind, explicitly given by the characteristic energy-dependent function:
\begin{equation}
    a_{\kappa}(E) = \frac{1}{\sqrt{E^{2}-4}} \left[ \left(\frac{E+\sqrt{E^{2}-4}}{2}\right)^{\kappa} - \left(\frac{E-\sqrt{E^{2}-4}}{2}\right)^{\kappa} \right].
\label{eq:chebyshev}
\end{equation}
From Eq.~(\ref{eq4}), we can directly invert the relation to express the intermediate amplitude $\psi_{m\kappa+1, n\kappa}$ as a linear combination of the neighboring vertex amplitudes:
\begin{equation}
    \psi_{m\kappa+1, n\kappa} = \frac{1}{a_{\kappa}} \psi_{(m+1)\kappa, n\kappa} + \frac{a_{\kappa-1}}{a_{\kappa}} \psi_{m\kappa, n\kappa}.
\end{equation}
By applying the transfer matrix along the corresponding spatial direction, similar expressions for $\psi_{m\kappa-1, n\kappa}$ and $\psi_{m\kappa, n\kappa\pm1}$ can be easily derived, which can be replaced by $\psi_{(m-1)\kappa,n\kappa}$ and $\psi_{m\kappa,(n\pm1)\kappa}$, respectively.

Substituting these recursive relations back into Eq.~(\ref{eq3}), we obtain a rigorously decimated eigenvalue equation that couples only the macroscopic vertex sites:
\begin{equation}
\begin{aligned}
    (a_{\kappa} E - 4 a_{\kappa-1}) \psi_{m\kappa, n\kappa} = & \psi_{(m+1)\kappa, n\kappa} + \psi_{(m-1)\kappa, n\kappa} \\
    & + \psi_{m\kappa, (n+1)\kappa} + \psi_{m\kappa, (n-1)\kappa} \\
    & + a_{\kappa} V_{m\kappa, n\kappa} \psi_{m\kappa, n\kappa}.
\end{aligned}
\label{eq:decimated_AA}
\end{equation}
To illuminate the physical equivalence and as shown in Fig.~\ref{fig1}(d), we can rescale the spatial coordinates by introducing the effective indices $m' = m\kappa$, $n' = n\kappa$, and $(m \pm 1)' = (m \pm 1)\kappa$, $(n \pm 1)' = (n \pm 1)\kappa$, and define an effective energy $E' = a_{\kappa} E - 4 a_{\kappa-1}$. Equation~(\ref{eq:decimated_AA}) then maps onto the 2D NH generalized AAH model:
\begin{equation}
    E' \psi_{m', n'} = \sum_{\langle m', n' \rangle} \psi_{m', n'} + V'_{m', n'} \psi_{m', n'},
\end{equation}
where the effective non-Hermitian quasiperiodic potential is amplified by the energy-dependent geometrical factor: $V'_{m', n'} = a_{\kappa} V_{m', n'}=2V_{\text{eff}}[\cos(2\pi\alpha_x m')+\cos(2\pi\alpha_y n')]$, where $V_{\text{eff}}=a_{\kappa} Ve^{i\theta}$.

For this 2D AAH model, the extended-localized transition point is rigorously dictated by using the dual transformation and
the critical condition is $|V_{\text{eff}}| = 1$. Thus, the exact Lyapunov exponent (LE) $\gamma(E)$—which dictates the exponential spatial decay of the wavefunction—can be analytically computed. LE establishes that for quasiperiodic systems, a strictly positive Lyapunov exponent $\gamma > 0$ unequivocally signals Anderson localized regime, $\gamma = 0$ indicate extended or mulifactor critical states. Translating the critical transition point $|V_{\text{eff}}| = 1$ back to the original parameters of our QDLL models, the exact MR equation reads:
\begin{equation}\label{eq10}
    |a_{\kappa}(E) Ve^{i\theta}| = 1.
\end{equation}
Because the eigenenergy $E$ is complex-valued, the modulus constraint in Eq.~(\ref{eq10}) implicitly defines a 1D closed curve in the $\text{Re}(E)$--$\text{Im}(E)$ plane. This continuous spectral boundary naturally generalizes the conventional real MEs into closed loops, formally establishing the concept of MRs. For these NH QDLLs, the extended states are strictly confined within the interior domain bounded by $|V_{\text{eff}}| < 1$, whereas the localized states occupy the exterior region corresponding to $|V_{\text{eff}}| > 1$.

\section{Analytical mobility rings and numerical results}\label{sec:4}

We can now analytically derive the MR boundaries for specific mosaic lattice configurations form the Eq.~(\ref{eq10}). For the simplest case $\kappa=1$, utilizing Eq.~(\ref{eq:chebyshev}), $a_1(E) = 1$, the system reduced the 2D  AAH model and the extended-localization transition point $|V| = 1$, and have no MRs. For the nontrivial decoration $\kappa=2$, we have $a_2(E) = E$, the critical boundaries $|E Ve^{i\theta}| = 1$: when $\theta=0$ corresponding to the Hermitian QDLL, the two MEs are $E_c = \pm 1/V$, when $\theta \in (0, \pi/2]$, the Hamiltonian ~(\ref{eq1}) is non-Hermitian and energy spectrum can form closed loops in the complex plane, the exact expression is 
\begin{equation}\label{eq11}
    {\rm{Re}}(E)^2+{\rm{Im}}(E)^2=1/V^2.
\end{equation}
Here the eigenenergy $E \in \mathbb{C}$, ${\rm{Re}}(E)$ and ${\rm{Im}}(E)$ represent the real and imaginary part of eigenenergy. As explicitly indicated by Eq.~(\ref{eq11}), provided that the complex quasiperiodic potential strength is non-vanishing ($V \neq 0$), the analytical boundary continuously traces out a single circular in the complex energy plane. 

To numerically verify the analytical results, we introduce the wavefunction fractal dimension (FD) as a macroscopic signature to characterize the localization transition. The FD is defined via the inverse participation ratio (IPR) as
$\Gamma = -\lim_{N \to \infty} \frac{\ln(\text{IPR})}{\ln N}$, where $\text{IPR} = \sum_{k,l} |\psi_{k,l}|^4$ and $N = (2\kappa - 1) \times L_x \times L_y$ denotes the total number of lattice sites. In the thermodynamic limit, $\Gamma \to 1$ characterizes an entirely extended state, whereas $\Gamma \to 0$ signifies a deeply localized state. We first examine the Hermitian limit ($\theta=0$) for the $\kappa=2$ modulation in Figs.~\ref{fig2}(a)-(c). By plotting $\Gamma$ for all eigenstates as a function of the potential strength $V$ and their corresponding real eigenvalues $E$, the analytical real MEs $|E_c| = 1/V$ (dashed red lines) are shown to cleanly separate the extended states ($\Gamma \approx 1$, yellow region) from the localized ones ($\Gamma \approx 0$, blue region). When non-Hermiticity is present ($\theta =\pi/2$), the real MEs generalize into a MR in the complex energy plane. According to Eq.~(\ref{eq10}), the critical condition for $\kappa=2$ reduces to $|E| = 1/V$. Because the eigenenergy $E$ is complex-valued, this relation implicitly defines a closed curve in the $\text{Re}(E)$--$\text{Im}(E)$ plane. In Figs.~\ref{fig2}(e) and \ref{fig2}(f), we map the complex energy spectra alongside the exact analytical MR boundaries derived from Eq.~(\ref{eq11}) (solid red curves) for $V=0.8$ and $V=1.5$, respectively. The theoretical boundaries precisely trace the sharp extended-localized phase transition: eigenstates encapsulated within the MR exhibit a robust extended nature ($\Gamma \approx 1$), whereas those expelled outside the MR are firmly localized ($\Gamma \approx 0$). Furthermore, the radius of the MR is intrinsically governed by an inverse dependence on $V$, shrinking as the quasiperiodic potential strength increases. To comprehensively validate our analytical derivations, we examine the complex energy spectra and the corresponding $\Gamma$ distributions across different non-Hermitian phase angles $\theta = \pi/4$ and $3\pi/4$ at $V=0.8$, as depicted in Figs.~\ref{fig3}(a)-(b). Although the spectral distribution in the complex plane depends sensitively on $\theta$, the analytical MR boundary derived from Eq.~(\ref{eq11}) remains invariant and consistently yields an accurate description of the localization transition, demonstrating the high reliability of our theoretical framework. Finally, the existence of the MRs is further corroborated by analyzing the real-space probability distributions of representative eigenstates. For $\theta = 3\pi/4$ and $V=0.8$, Figs.~\ref{fig3}(c)-(d) illustrate the spatial profiles $|\psi|^2$ of an extended eigenstate inside the MR ($E=-0.9750 - 0.2198i$) and a localized eigenstate outside the MR ($E=3.2132 - 1.4185i$), respectively, exhibiting clear delocalized and tightly bound features in complete agreement with our theoretical predictions.

\begin{figure*}[t]
	\centering
	\includegraphics[width=0.90\textwidth]{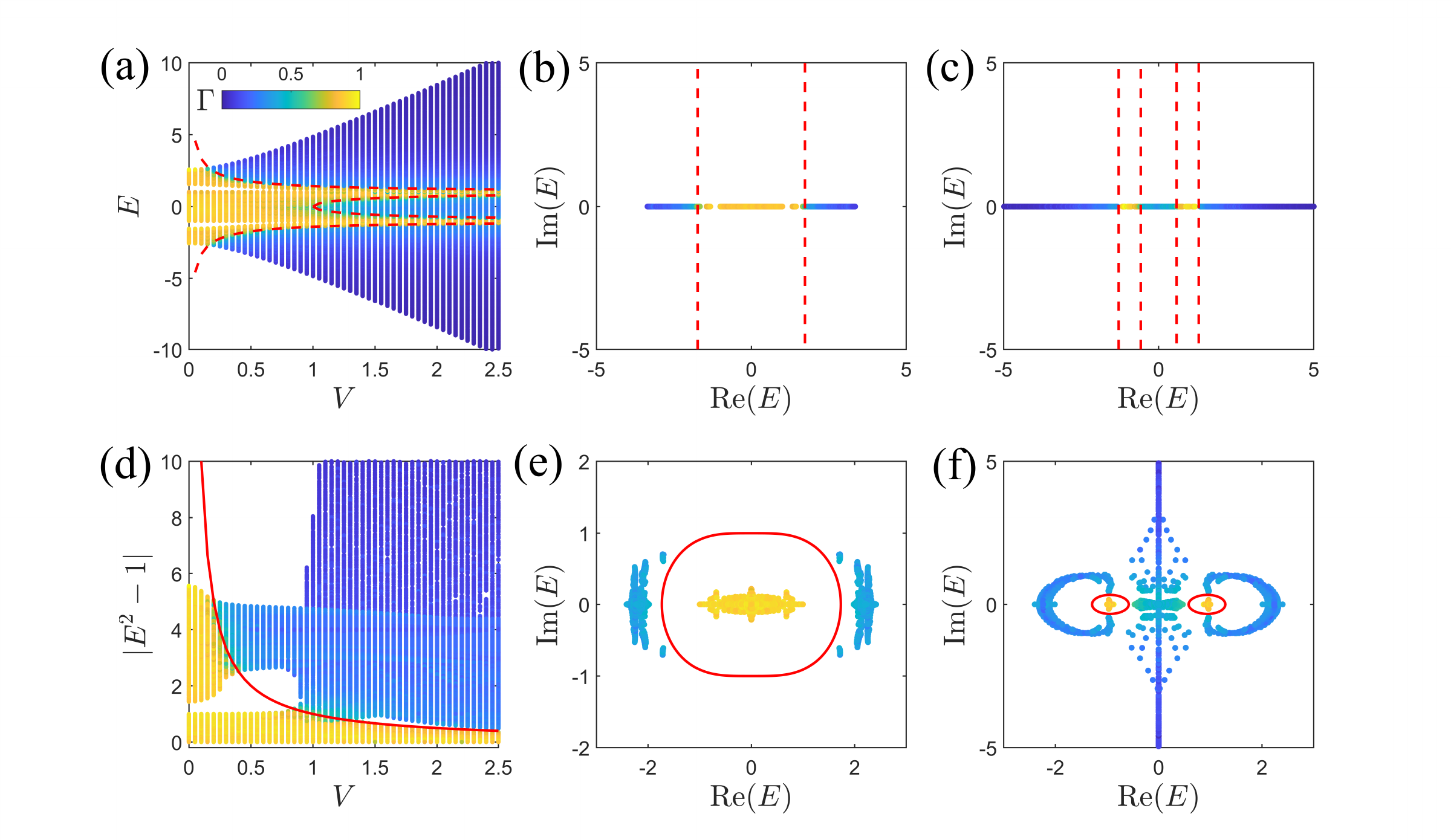}
        \caption{Energy spectrum, mobility edges (MEs), and mobility rings (MRs) for the $\kappa = 3$ case. (a) Wavefunction fractal dimension $\Gamma$ as a function of the quasiperiodic potential strength $V$ and real eigenvalues $E$ in the Hermitian limit ($\theta = 0$). The red dashed lines denote the critical ME boundary $E_c = \pm \sqrt{1 \pm 1/V}$, which separating from extended to localized states. (b), (c) Spectral distributions for $V = 0.5$ and $V = 1.5$ at $\theta = 0$. (d) Complex energy spectrum and exact MR boundary $|E^2 - 1| = 1/V$ (solid red curve) for the non-Hermitian case $\theta = \pi/2$. (e), (f) Spectral phase diagrams in the complex energy plane for $V = 0.5$ and $V = 1.5$ at $\theta = \pi/2$. The eigenstates within the MR are extended ($\Gamma \approx 1$), while the eigenstates outside the MR are localized ($\Gamma \approx 0$). As $V$ increases beyond the critical threshold $V_c=1$, the MR undergoes a topological phase transition—splitting via a critical Lemniscate of Bernoulli—into two independent, decoupled sub-rings. Here, $t = 1$ is set as the unit of energy, and the system size are $L_x = L_y = 55$ under periodic boundary conditions.}
	\label{fig4}
\end{figure*}

\begin{figure}[t!]
    \hspace{-1.0cm}
	\centering
	\includegraphics[width=0.50\textwidth]{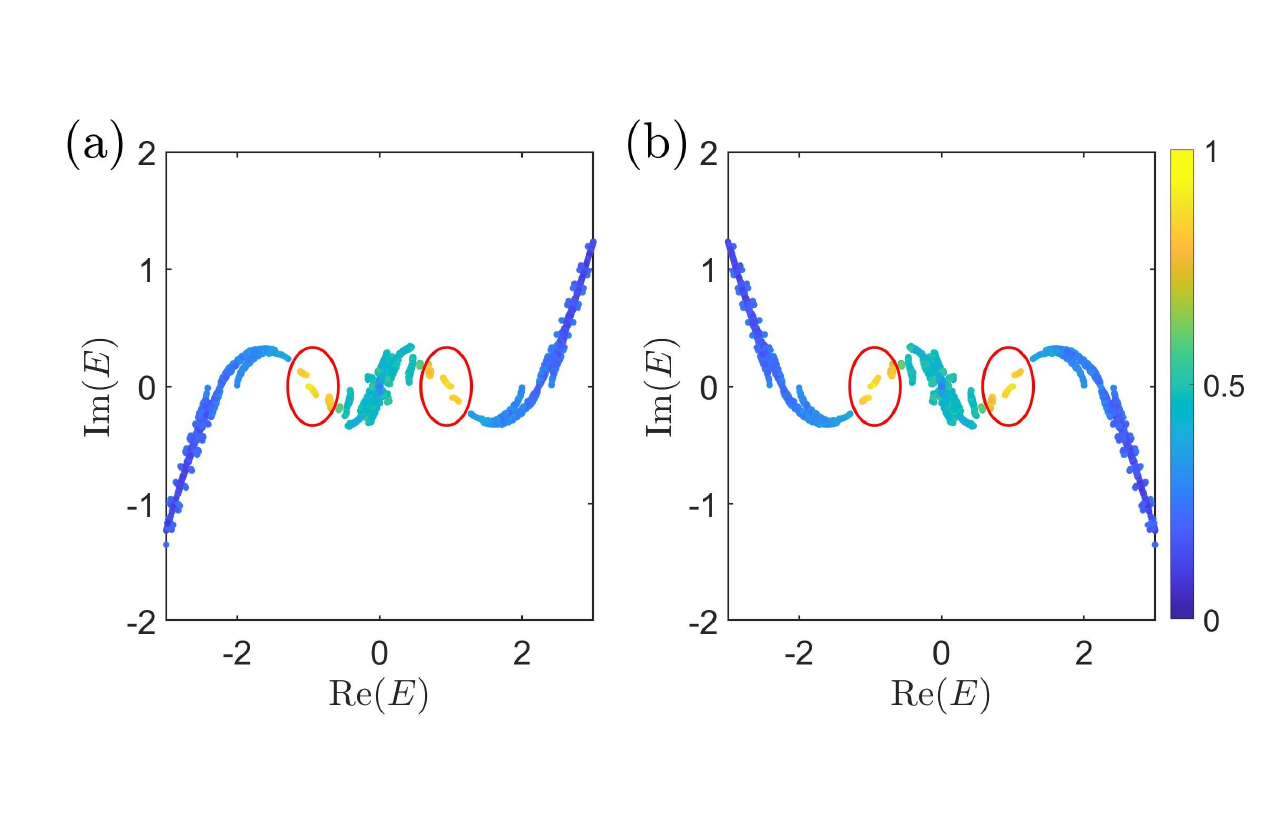}
        \caption{The MRs for different $\theta$ for $\kappa=3$ case. (a)-(b) represent the  $\theta=\pi/4$  and $\theta=3\pi/4$. Although the complex spectral features shift markedly with varying $\theta$, the analytical sub-ring boundaries defined by Eq.~(\ref{eq12}) remain invariant.}
    \hspace{-1.0cm}
    \label{fig5}
\end{figure}

For the $\kappa=3$ case, the characteristic polynomial generalizes to $a_3(E) = E^2 - 1$. In the Hermitian limit ($\theta = 0$), the critical boundary $E_c = \pm \sqrt{1 \pm 1/V}$ [dashed red lines in Fig.~\ref{fig4}(a)], and the system exhibits two or four distinct real MEs respectably for $V\leq1$ and $V>1$. We set $V=0.5$ and $V=1.5$, which are directly confirmed by the vertical phase boundaries in the real-energy spectrum Figs.~\ref{fig4}(b)-(c). When non-Hermiticity is introduced ($\theta =\pi/2$), the localization boundary condition $|a_3(E)V| = 1$ simplifies to $|E^2 - 1| = 1/V$. Expressed in Cartesian coordinates $E = \text{Re}(E) + i\text{Im}(E)$, the exact MR equation in the complex energy plane reads
\begin{equation}\label{eq12}
\left[\text{Re}(E)^2 + \text{Im}(E)^2\right]^2 - 2\left[\text{Re}(E)^2 - \text{Im}(E)^2\right] = \frac{1}{V^2} - 1.
\end{equation}
In Fig.~\ref{fig4}(d), we plot the effective modulus $|E^2 - 1|$ of the eigenenergy against the modulation strength $V$. The analytical curve $1/V$ (solid red line) establishes a sharp demarcation line: the extended phase ($\Gamma \approx 1$, yellow region) lies entirely below the curve, whereas the localized phase ($\Gamma \approx 0$, blue region) occupies the domain above it. Equation~(\ref{eq12}) describes a family of Cassini ovals in the complex energy plane, whose geometric topology undergoes a profound dynamical evolution driven by $V$. For weak quasiperiodic potential strengths ($V < 1$), the critical boundary forms a single integrated loop enclosing both foci $E = \pm 1$, as illustrated by the continuous red curve in Fig.~\ref{fig4}(e) for $V=0.5$. As $V$ increases beyond the critical threshold $V_c=1$, the MR undergoes a topological phase transition—splitting via a critical Lemniscate of Bernoulli—into two independent, decoupled sub-rings centered around $E = \pm 1$. Figure~\ref{fig4}(f) illustrates this two-ring topology for $V=1.5$, where the analytical sub-rings (solid red loops) perfectly encapsulate two disjoint clusters of extended states ($\Gamma \approx 1$). To further verify the universality of this algebraic boundary, we examine the complex energy spectra under different non-Hermitian phase angles $\theta = \pi/4$ and $3\pi/4$ at $V=1.5$ in Figs.~\ref{fig5}(a) and \ref{fig5}(b). Although the complex spectral features shift markedly with varying $\theta$, the analytical sub-ring boundaries defined by Eq.~(\ref{eq12}) remain strictly invariant and continuously yield an accurate description of the extended-localized transition. These numerical results rigorously confirm the existence and rich topological evolution of mobility rings in two-dimensional non-Hermitian systems.

\section{Conclusion}\label{sec:5}

In summary, we have systematically investigated the Anderson localization transitions and derived the exact MRs in a class of 2D QDLL models. By exploiting the discrete spatial symmetries inherent to these decorated geometries, we exact  mapped the 2D multi-sublattice Hamiltonians onto effective 2D generalized AAH models. This mapping allowed us to derive explicit, closed-form critical conditions Eq.~(\ref{eq10}) governing the exact MR boundaries in the complex energy plane, circumventing the formidable convergence bottlenecks typical of high-dimensional transfer matrix calculations. Our analytical predictions are in quantitative agreement with numerical evaluations of wavefunction fractal dimensions and real-space spatial profiles. Crucially, we unveiled a topology-driven evolution of the MR geometry determined by the decoration parameter $\kappa$: while $\kappa=2$ systems feature a single circular MR, $\kappa=3$ systems exhibit a dynamical transition where a single Cassini oval splits into two decoupled sub-rings as the potential strength crosses $V_c=1$. Furthermore, we demonstrated that while non-Hermitian phase $\theta$ variations alter the complex spectral distributions, the exact MR boundary remains fundamentally invariant. By providing a rare exact benchmark in two dimensions, our work opens new avenues for exploring higher-dimensional non-Hermitian quasicrystals, localization transitions, and quantum transport in open dissipative architectures.

\section*{Acknowledgments}

This work is supported by the National Natural Science Foundation of China (Grants No.~12304388, No.~12304290, and No.~12505017), the Beijing National Laboratory for Condensed Matter Physics (Grant No.~2025BNLCMPKF017), and the Fundamental Research Funds for the Central Universities.

\bibliography{Localization}
\end{document}